\documentclass[12pt,preprint]{aastex}

\begin{document}

\title{Spitzer/IRS Imaging and Spectroscopy of the luminous infrared
  galaxy NGC\,6052 (Mrk 297)}

\author{D. G. Whelan\altaffilmark{1}, D. Devost\altaffilmark{1},
  V. Charmandaris\altaffilmark{2,3}, J.A. Marshall\altaffilmark{1},
  J. R. Houck\altaffilmark{1}}

\altaffiltext{1}{Center for Radiophysics and Space Research, Cornell
  University, 219 Space Sciences Building, Ithaca, NY 14853-6801, USA}

\altaffiltext{2}{University of Crete, Department of Physics, P.O. Box
  2208 GR-71003, Heraklion, Greece}

\altaffiltext{3}{IESL, Foundation for Research and Technology-Hellas,
  GR-71110, Heraklion, Greece, and Chercheur Associ\'{e}, Observatoire
  de Paris, F-75014, Paris, France}

\email{whelan@isc.astro.cornell.edu,
  devost@astro.cornell.edu,\\vassilis@physics.uoc.gr,
  jam258@cornell.edu, jrh13@cornell.edu}

\begin{abstract}
  We present photometric and spectroscopic data of the interacting
  starburst galaxy NGC\,6052 obtained with the Spitzer Space
  Telescope. The mid-infrared (MIR) spectra of the three brightest
  spatially resolved regions in the galaxy are remarkably similar and
  are consistent with dust emission from young nearly coeval stellar
  populations. Analysis of the brightest infrared region of the
  system, which contributes $\sim$18.5\% of the total 16\micron\ flux,
  indicates that unlike similar off-nuclear infrared-bright regions
  found in Arp\,299 or NGC\,4038/9, its MIR spectrum is inconsistent
  with an enshrouded hot dust (T$>$300K) component. Instead, the three
  brightest MIR regions all display dust continua of temperatures less
  than $\sim$ 200K. These low dust temperatures indicate the dust is
  likely in the form of a patchy screen of relatively cold material
  situated along the line of sight. We also find that emission from
  polycyclic aromatic hydrocarbons (PAHs) and the forbidden atomic
  lines is very similar for each region. We conclude that the
  ionization regions are self-similar and come from young ($\lesssim$
  6 Myr) stellar populations. A fourth region, for which we have no
  MIR spectra, exhibits MIR emission similar to tidal tail features in
  other interacting galaxies.
\end{abstract}


\section{Introduction}
\label{int}

NGC\,6052 (also known as Mrk\,297, Arp\,209, VV\,86) is an interacting
system located at a distance of 70.4\,Mpc and having an L$_{\rm
  IR}=1\times10^{11}~$L$_{\odot}$ \citep{Sanders03}, which puts it
into the class of luminous infrared galaxies (LIRGs). The peculiar
morphology of the system has made it the subject of several
multi-wavelength observations
\citep[e.g.][]{Weedman72,Taniguchi91}. There is a general consensus
that NGC\,6052 is a merger of two galaxies
\citep{Burenkov88,Alloin79,Taniguchi91}, although the exact
morphological type of the galaxies involved is still
unclear. H-$\alpha$ imaging revealed a number of star formation
regions throughout the system \citep{Cairos01II}. \citet{Hecquet87}
identified the locations of forty-three different optical knots in
NGC\,6052, distributed along the central bar-like structure of the
galaxy and on the spiral-arm/tidal feature which extends along the
north-south direction.

An extensive study of the infrared properties of the system was
performed with the Infrared Space Observatory (ISO) involving both MIR
imaging with ISOCAM and spectroscopy with PHOT-S and LWS
\citep{Metcalfe05}. Seven MIR knots were visible in the 14.3\micron\
images of the system. The most luminous of these sources, identified
as ``region-1'' by \citet{Metcalfe05}, is approximately 8\arcsec\ to
the north of the main bulge of the galaxy. This source contributes
$\sim$22\% of the 15\micron\ flux of NGC\,6052, yet the ISO data
indicated that it was faint optically.  This suggested that it may be
a more extreme example of the infrared-bright, optically-obscured
``overlap region'' region found between the two galactic nuclei of the
Antennae galaxies \citep{Vigroux96}. The MIR luminosity of the
NGC\,6052 source is a factor of $\sim$14 higher than that of the
Antennae source. An alternative possibility is that the source is an
obscured active galactic nucleus, like IC\,694 \citep{Gallais04}. This
hypothesis was tested by \citet{Metcalfe05} using infrared
diagnostics, who concluded that an AGN in the NGC\,6052 source would be
deeply embedded and needed to be confirmed with hard X-ray
observations.

In this paper, we present photometry and spectroscopy of the three
brightest MIR regions in NGC\,6052 and photometry of a fourth region,
using the Infrared Spectrograph\footnote{The IRS was a collaborative
  venture between Cornell University and Ball Aerospace Corporation
  funded by NASA through the Jet Propulsion Laboratory and the Ames
  Research Center.}  (IRS) on board the Spitzer Space Telescope. We
present the observations and data reduction in \S 2, the analysis in
\S 3, and discuss the implications of our observations in \S 4.

\section{Observations and Data Reduction}
\label{sec:Observations}

\begin{figure}
\epsscale{1.0}
\plotone{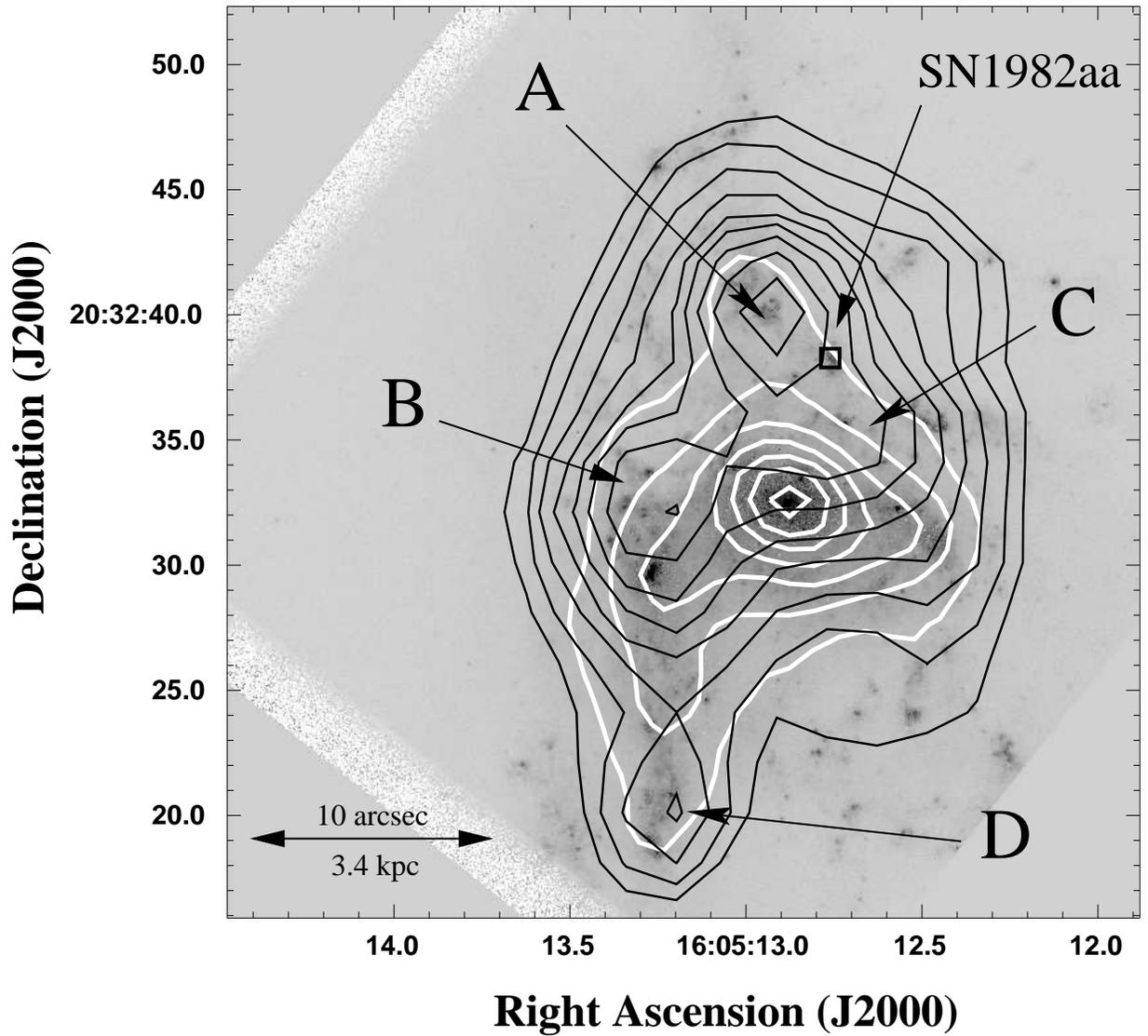}
\caption{The HST optical (F 555W) image of the galaxy, over-plotted
  with 2MASS K-band image (white) and Spitzer/IRS 16\micron\ image
  (black). The three regions from which IRS spectra were extracted are
  labeled A, B, and C. Source D is also labeled. Note that region A
  emits most at 16 \micron. The position of SN 1982aa (see
  \S\ref{sec:RadioContinuumEmission}) is marked by a
  square.\label{puimage}}
\end{figure}

NGC\,6052 was observed in mapping mode with the Spitzer/IRS
\citep{Werner04,Houck04} in both the short-low (SL, 5.2-14.5\micron)
and long-low (LL, 14.0-38.0\micron) modules. The SL data were obtained
on 15 July 2004 and the LL data on 4 March and 17 July 2004, resulting
in a map of the central area of the system over the whole
5.3--37\micron\ range. The total exposure time per IRS slit was 44~sec
for SL, and 20~sec for LL. Target acquisition was performed using the
IRS blue (16 \micron) peak-up camera.\footnote{For details on the IRS
  peak-up, see chapter 7 of Spitzer Observing Manual at
  http://ssc.spitzer.caltech.edu/documents/som/} As a result of the
IRS peak-up procedure, a 16\micron\ image with a total exposure time
of 201~sec centered on the brightest position of our target was
obtained. The field of view of the image is
60\arcsec$\times$72\arcsec\ and its scale is 1.8\arcsec\
pixel$^{-1}$. All the observations were processed by the {\em Spitzer}
Science Center pipeline version 13.2. The stability of Spitzer's
pointing enables the registration of the image on a world coordinate
system (WCS) grid with an accuracy which is better than the pixel size
of the peak-up array. The 16\micron\ contours along with 2MASS
contours are overlaid on the HST image in Figure \ref{puimage}.

Two different spectral extractions were performed. First, an
integrated spectrum of NGC\,6052 was extracted using the Spectral
Modeling, Analysis, and Reduction Tool \citep[SMART
Ver. 5.5.1][]{Higdon04} for all the SL and LL pointings of the
spectral map. The 2D images were coadded together and a standard full
slit extraction was performed on each coadded low resolution
aperture. The resulting spectrum is shown in Figure
\ref{intspec}. Second, we extracted individual sources (labeled A, B,
and C in Figure \ref{puimage}) in the SL module. Sources A and C were
located within the same slit, requiring the development of custom
software to perform a Gaussian model fit to the spatial profile of
multiple sources within a slit. Each source in the slit was traced
using information about the location of each wavelength element on the
array \citep{Whelan05}. We estimate the cross-contamination between
sources A and C to be $\lesssim\ 5$\%.  The full-width at half-maximum
of each curve fit was assumed to be the same as that of a stellar
point-source, and therefore was allowed to vary with
wavelength. Source B was extracted in a likewise fashion, for one
source within a slit. The spectral map does not cover source D.

We flux calibrated the data using observations of $\alpha$ Lac that
were extracted in exactly the same manner as the object spectra. The
extracted spectrum from the star was used in conjunction with a
template to produce a Relative Spectral Response Function (RSRF)
\citep{Cohen03}. The RSRF was applied to each extracted spectrum of
NGC\,6052. The spectra of sources A, B and C are shown in
Figure~\ref{multispec}. We estimate our spectrophotometry to be better
than 10\%.

The flux from the various polycyclic aromatic hydrocarbon (PAH)
features was calculated using the fitted Drude profiles from the
spectral decompositions described in
\S\ref{sec:SpectralDecomposition}. The [Ne~{\sc ii}]12.81\micron\ line
flux was also measured using this fitting method since it is blended
with the 12.7\micron\ PAH feature in the SL module. All other atomic
lines have a well defined continuum, so we calculated their total
fluxes by simply summing the pixel values following a local continuum
subtraction. Our PAH feature and line measurements are presented in
Table~\ref{tabflux}.

Near-IR J, H, and K aperture photometry was performed on the 2MASS
images \citep{Skrutskie06} at the positions of sources A, B, C, and D.
These measurements are included with the spectra in
Figure~\ref{multispec}, as they are important to the spectral
decompositions (discussed in \S\ref{sec:SpectralDecomposition}). To
compare the overall optical morphology of the system to the thermal
dust emission we use a V-band (F555W) Hubble Space Telescope (HST)
image of the galaxy from the HST archive (Figure~\ref{puimage}).

We also use archival broadband MIR images of NGC\,6052 obtained with
the Infrared Array Camera (IRAC) instrument on Spitzer
\citep{Fazio04}. The images were obtained on March 27, 2005 with the
3.6, 4.5, 5.8, and 8.0 \micron\ broadband filters and they were
processed with the Spitzer Science Center pipeline version 14.0. The
FWHM of the point spread function of the images varies between
1\arcsec.6 and 2\arcsec.0 for the different filters. The images have a
pixel size of 1.2\arcsec\ $\times$1.2\arcsec\ and a field of view
5.2\arcmin\ $\times$ 5.2\arcmin\. The total integration time per
filter was 150s.

\section{Analysis}

\subsection{Global Morphology}
\label{sec:GlobalMorphology}

The distribution of dust and star emission from NGC\,6052 is displayed
in Figure~\ref{puimage}, where we overlay contours of the 16 \micron\
Blue Peak-Up (BPU) and 2MASS K-band images onto the optical HST
image. The regions marked A, B, C, and D align with optically visible
HII regions, but none are associated with the two brightest optical
regions, which are labeled by \citet{Alloin79} as the compact regions
of the merged galaxies forming NGC\,6052. See Table~\ref{srcid} for a
comparison of our source identification to the MIR regions from
\citet{Metcalfe05} and the B-R color knots from \citet{Hecquet87}.

\begin{figure}
\epsscale{1.0}
\plotone{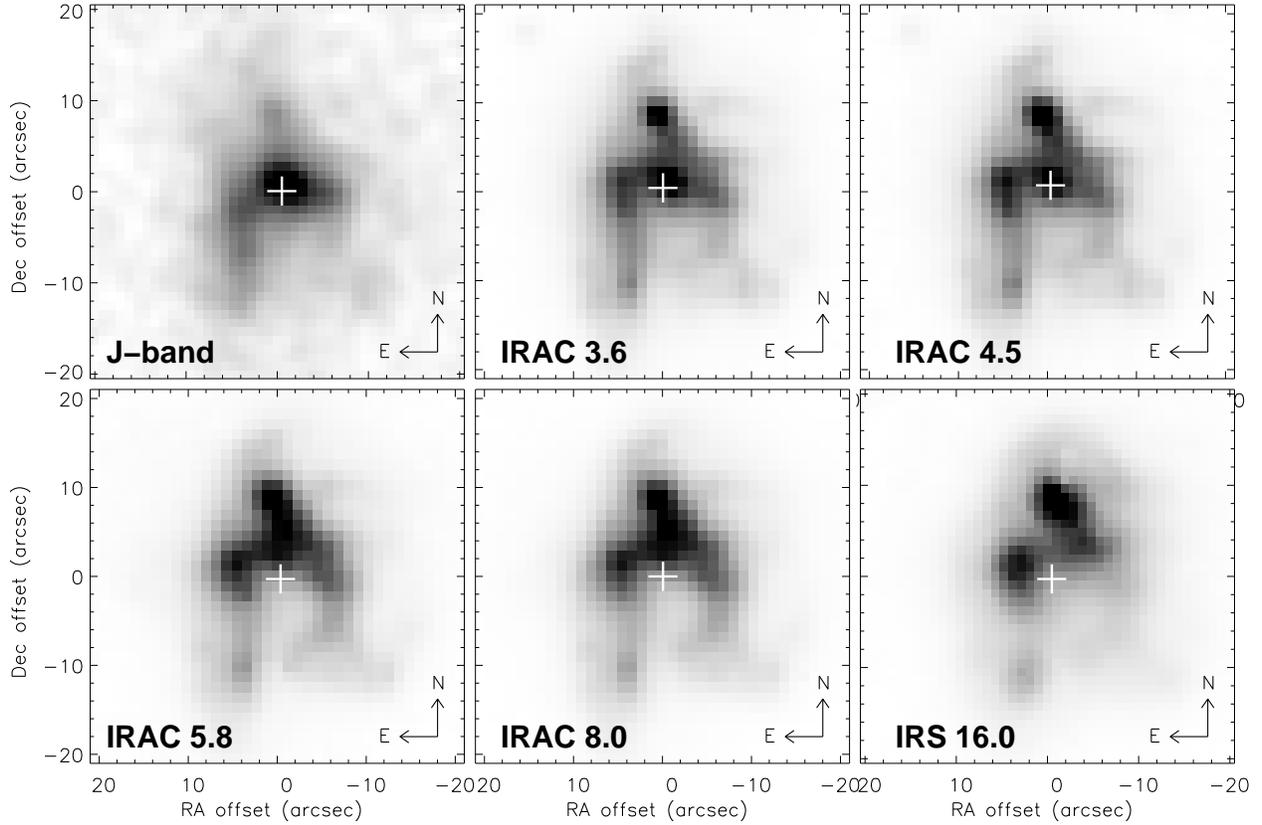}
\caption{A snapshot composite of the near-IR 2MASS J-band image of
  NGC\,6052, along with the 3.5, 4.0, 5.8, and 8.0\micron\ IRAC images
  as well as the IRS 16\micron\ peak-up image. Each snapshot is
  centered at RA(J2000)=16:05:12.89, Dec(J2000)=+20:32:32.5, which
  coincides with the location of the near-IR bulge of the galaxy as
  indicated by the J-band image.\label{astrometry}}
\end{figure}

\begin{figure}
\epsscale{0.6}
\plotone{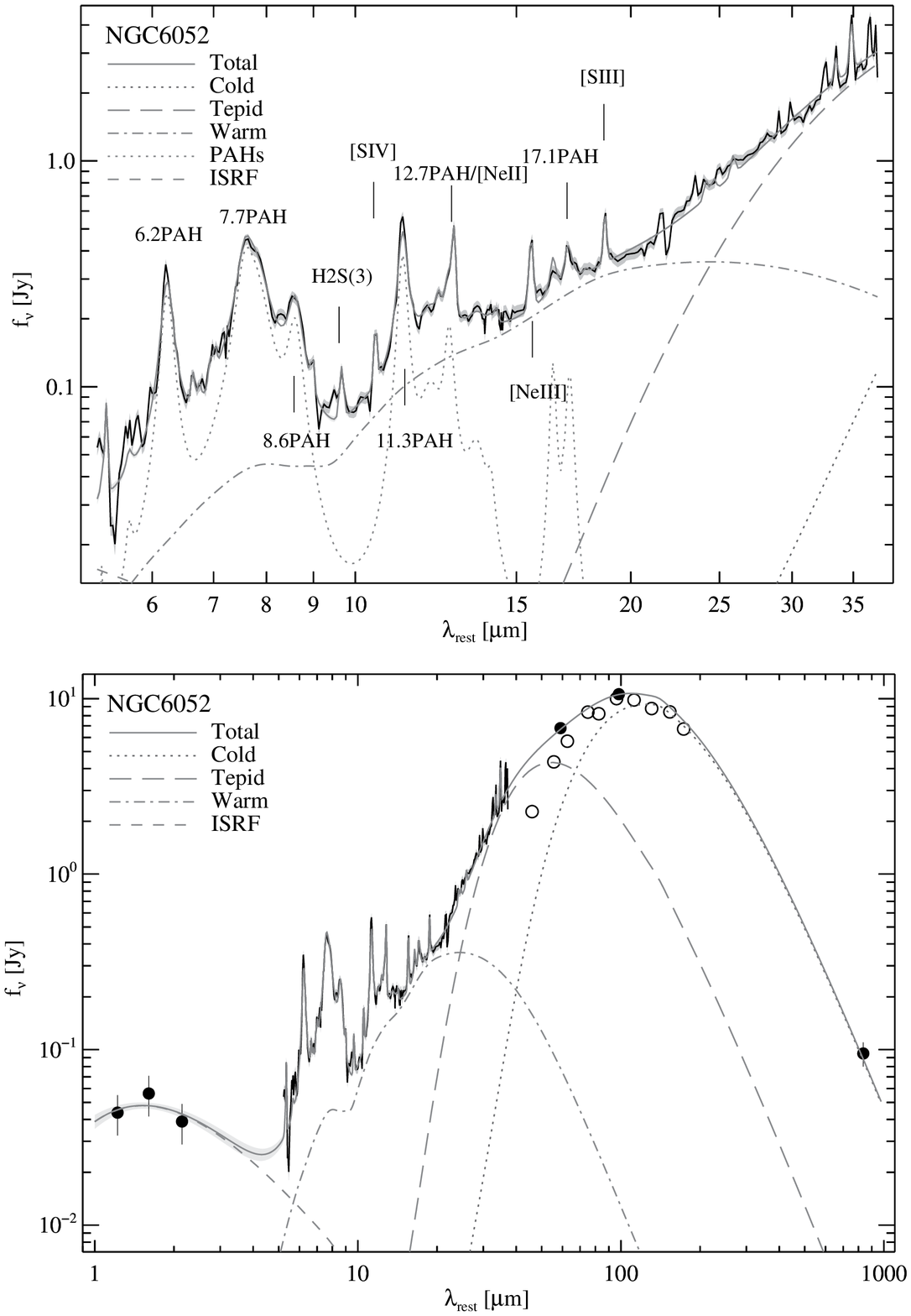}
\caption{Top: The integrated Spitzer/IRS spectrum of NGC\,6052, with
  all strong MIR spectral features marked. Bottom: The 1 \micron\ to
  100 \micron\ SED of the system. The far-IR photometric points for
  the systems were obtained from \citet{Metcalfe05} and \citet{Dunne00}
  (filled symbols were used in the spectral decomposition while open
  symbols were not), and the near-IR photometric points are from
  \citet{Spinoglio95}. The warm dust, PAH, and interstellar radiation
  field (ISRF) components of the model fits are also
  indicated.\label{intspec}}
\end{figure}

\begin{figure}
\centering
\includegraphics[width=0.5\textwidth]{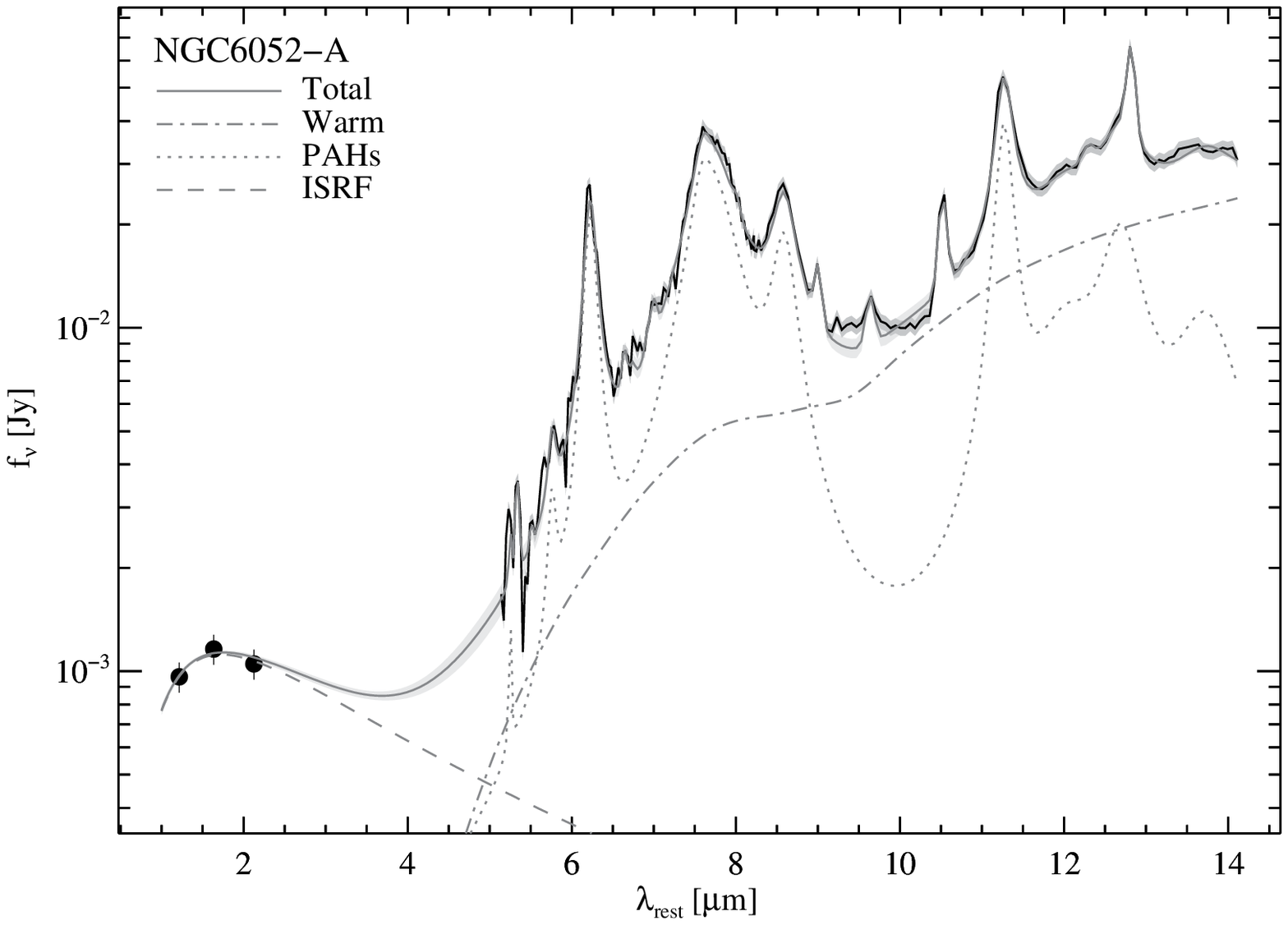}
\includegraphics[width=0.5\textwidth]{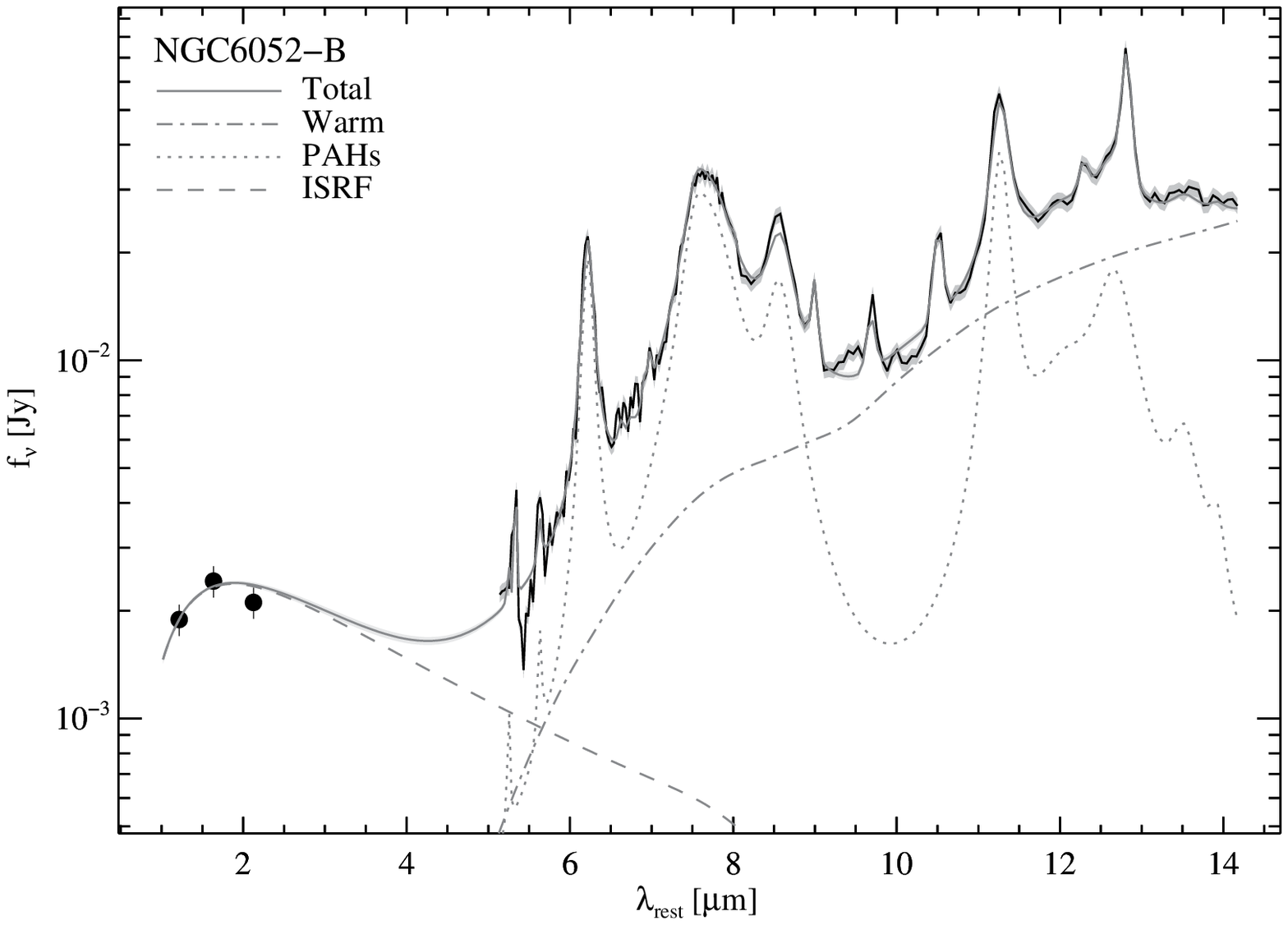}
\includegraphics[width=0.5\textwidth]{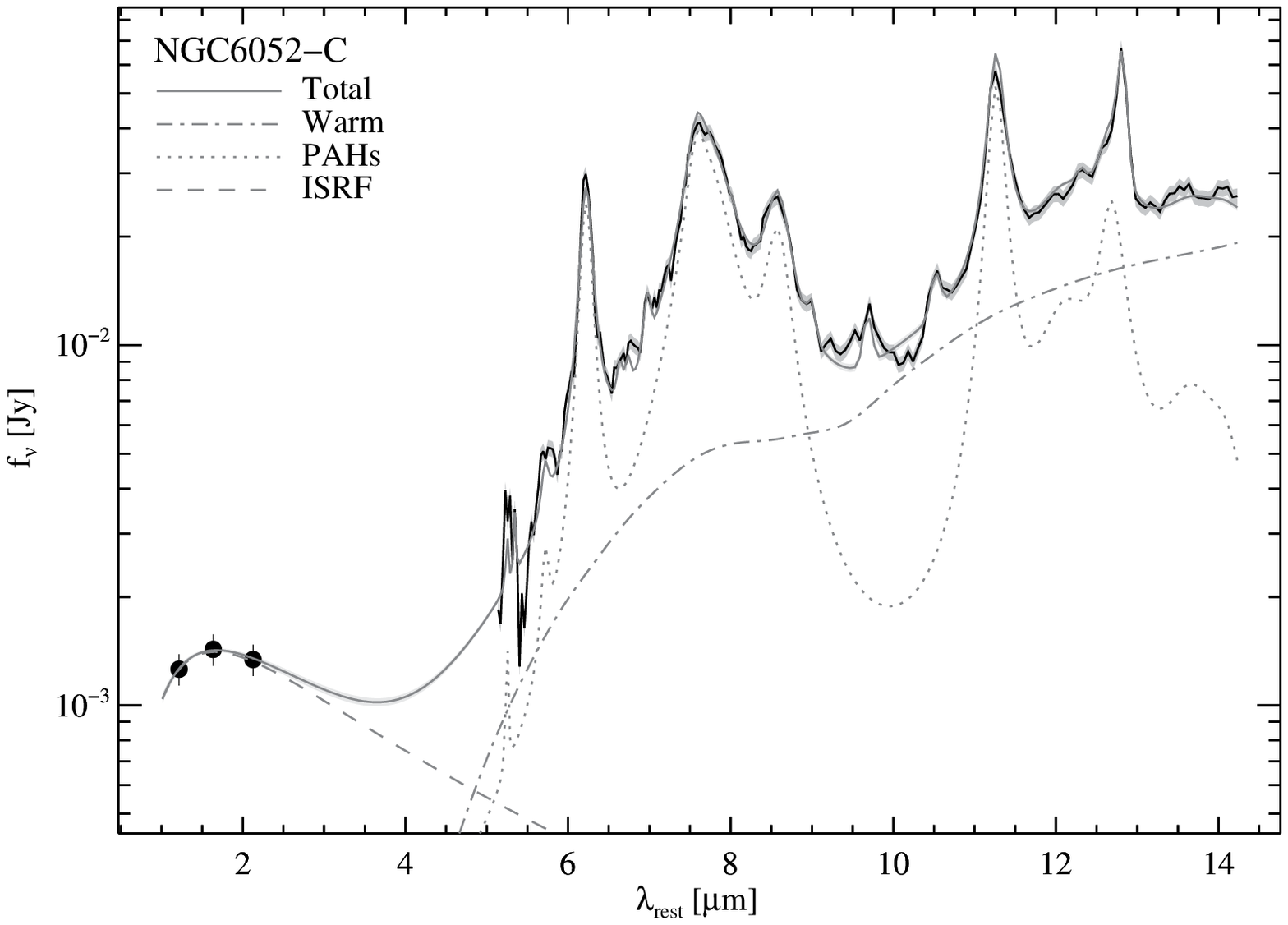}
\caption{The $\sim$5--15\micron\ IRS SL spectrum of sources A, B, and C. The
  filled circles indicate the J,H, and K 2MASS points for each region. The
  warm dust, PAH, and interstellar radiation field (ISRF) components of the
  model fits are also indicated.\label{multispec}}
\end{figure}

Source alignment was based on the reconstructed pointing of Spitzer
and the astrometry of the 2MASS image. With the better registration of
Spitzer and 2MASS, we find that source A is shifted 4\arcsec\ north
relative to the optical, compared to the ISO data presented in
\citet{Metcalfe05}, and is not as optically-obscured as was previously
thought based on ISOCAM data. The pointing accuracy of Spitzer is a
factor of two smaller than the 1.8\arcsec\ pixel size of the
16\micron\ image. We confirmed our astrometry using 2MASS J, IRAC 3.6,
4.5, 5.8 and 8.0 \micron, and IRS 16.0 \micron\ images, as shown in
the Figure~\ref{astrometry} mosaic. This mosaic best illustrates the
centroid shifts of the sources. The location of the near-IR bulge of
the galaxy is indicated by a white cross and we observe an obvious
shift in the emission as wavelebgth changes from predominantly main
sequence star light in the 2MASS image, to PAH and hot dust in the
IRAC bands, to just dust at 16.0 \micron. It is clear that the
concentration of warm dust and PAH emission as traced by the IRAC 5.8
and 8\micron\ bands lies to the north and east of the NIR bulge, in
regions A, B and C, and more faintly along the eastern tidal arm,
where source D lies. At 16\micron, where more embedded star forming
regions may be probed, the spatial resolution is lower, but most of
the emission originates from regions A and B.

The total MIR flux of the galaxy at 16\micron\ is measured as 260 mJy
$\pm$ $\sim$ 12\% (see Table~\ref{srcid} for a complete list of flux
densities). \citet{Metcalfe05} found a 15 \micron\ flux density of 289
mJy $\pm$ $\sim$ 15\%. This flux density, as well as our 16\micron\ image,
globally match the MIR morphology presented by \citet{Metcalfe05}
using ISOCAM.

\citet{Metcalfe05} showed that region A in NGC\,6052 contributes 22\%
of the total emission at 15 \micron. In contrast, see
Table~\ref{srcid}, we find that region A contributes only 18.5\% of
the total emission at 16 \micron. This difference in flux density
between the ISOCAM measurements and the Spitzer IRS BPU measurements
is equal to 18.5 mJy. A similar difference of 67 mJy between the
ISOCAM 7.7 \micron\ and the IRAC 8 \micron\ fluxes, where the ISOCAM
measurement is again higher than what we measure, is also noted. The
reason for these differences is unclear. All of our flux measurements
appear to be systematically lower than those given by
\citet{Metcalfe05}.

\subsection{Radio Continuum Emission}
\label{sec:RadioContinuumEmission}

Since radio emission is also an extinction-free tracer of star forming
regions, we use the high resolution radio continuum map of
\citet{Yin94} to compare the radio morphology of the galaxy to our
16\micron\ image. The radio map gives evidence for a bright radio
supernova located at the position of knot 14 in \citet{Hecquet87}
\citep{Lonsdale92, Sage93, Yin94}. Knot 14, which is unresolved in the
MIR, does not coincide with either regions A or C (see
Figure~\ref{puimage}), but it does fall within the SL slit. As a
result it is possible that some of the emission from the source
surrounding the radio supernova may contribute to the extracted IRS
spectra of sources A and C. \citet{Yin94} notes that the flux from
Hecquet's knot 14 is either from SN 1982aa (five years after the
explosion) or from an HII region. As is discussed by
\citet{Metcalfe05} the most likely type for SN 1982aa is Ib/c. The HST
image of Figure~\ref{puimage} indicates that the position attributed
to knot 14 of \citet{Hecquet87} is a diffuse HII region. However, the
supernova presumed to be associated with knot 14 is not visible in
recent optical and near-IR images and the spectra of regions A and C
do not greatly differ in spectral continuum slope or feature strength
from that of region B (see \S\ref{sec:SpitzerSpectroscopy} below). In
other words, we see no MIR features associated with the supernova, and
its HII region is unresolved.

There is a region to the northeast of SN 1982aa in Yin's radio
continuum map which is associated with our region A. There is also
diffuse radio emission associated with regions B and C. Region D is
not resolved in the radio continuum map.

\subsection{Spitzer Mid-IR Spectroscopy}
\label{sec:SpitzerSpectroscopy}

Our spectrum, displayed in Figure~\ref{intspec}, is in overall
agreement with the 6--11\micron\ spectrum obtained by ISO/PHOT-S in
\citet{Metcalfe05}. Our spectrum displays strong emission from
polycyclic aromatic hydrocarbons (PAHs) at 6.2, 7.7, 8.6, 11.25 12.7
and 17.1 \micron. Several fine structure lines, including [Ne~{\sc
  ii}], [Ar~{\sc iii}], [S~{\sc iv}], and [Ne~{\sc iii}], are also
clearly visible in the spectrum, as well as the molecular hydrogen
lines: S(3)$\lambda$9.66\micron; and S(2)$\lambda$12.27\micron. The
ratio [Ne~{\sc iii}]/[Ne~{\sc ii}]=0.84 $\pm$0.16 while the [S~{\sc
  iv}]/[S~{\sc iii}]=0.52 $\pm$0.18. This is similar to typical values
( $\sim$1 and $\sim$0.5 respectively) found in other starburst
galaxies \citep{Brandl06} and lower than those seen in Blue Compact
Dwarf galaxies \citep{Wu06}.

In addition, the narrow width of the IRS SL slits (3.6\arcsec) enable
us to obtain 5.3--15\micron\ spectra for the three brightest MIR
regions in the galaxy. Extending the spectrum of those regions to
longer wavelengths was not possible since the individual regions are
unresolved within the 9.7$\arcsec$ width of the IRS 15--37\micron\ LL
slit. Inspection of our IRS spectra, displayed in
Figure~\ref{multispec}, reveals that the MIR spectra for all three
regions are surprisingly similar. For instance, even though region A
is brighter than regions B or C at 16 \micron, lending to a marginally
steeper MIR slope, the PAH spectrum shows few noticeable differences.
Of special note is the shallow 9.6\micron\ silicate absorption
feature. We will elaborate on a possible physical explanation of these
features in \S\ref{sec:Discussion}.

Table~\ref{tabflux} lists the integrated line fluxes and major
PAH features for the whole system, as well as in the individual
regions A, B, and C.

\subsection{Spectral Decomposition}
\label{sec:SpectralDecomposition}

In order to characterize the properties of the emitting and obscuring
dust (including dust temperatures, optical depths, and PAH feature
strengths), we decompose the infrared spectral energy distribution of
each spectrum into emission from: (1) multiple dust components at
different characteristic temperatures; (2) stellar photospheres; (3)
PAHs; and (4) atomic and molecular lines. The contribution from each
dust component is calculated using a realistic dust model consisting
of a distribution of thermally emitting carbonaceous and silicate
grains. The model accounts for stochastic emission from PAHs by
fitting observationally derived PAH templates to the spectra. Details
of the method are described in \citet{Marshall07} (and see
\citet{Armus06, Armus07} for additional applications).

For the integrated spectrum of the galaxy we find that three dust
components fit the data well, at temperatures of 25, 57, and 203
$\pm$5 K. The $\tau$$(9.7)$ for the global spectrum was found to be
0.9 $\pm$ 0.5, where the uncertainty is due to statistics, as well as
the dependence on model choice for the PAH template and dust geometry.
This number is far more than the optical depth computed by
\citet{Takagi03} of $\tau$$_{V}$ $=$ 0.9 which corresponds to
$\tau$$(9.7)$ $\sim$0.1 using a standard extinction curve.
\citet{Takagi03} derive their $\tau$$_{V}$ from the modeled mass
fraction of hydrogen and the mass fraction of stars. In contrast, our
optical depth estimate is derived from the MIR spectrum.

The integrated PAH luminosity of the system is $\sim$6.5\% $\pm$ 0.5\%
of the total L$_{\rm IR}$, nearly 50\% higher than the value found for
the prototypical starburst galaxy NGC\,7714 \citep{Brandl04,
  Marshall07}. The sum of the 6.2\micron\ PAH feature strengths from
regions A, B, and C is $\sim$19\% of the total 6.2\micron\ PAH feature
strength from the integrated spectrum. In comparison, the 14\micron\
flux from regions A, B, and C totals to $\sim$40\% of the 14\micron\
flux from the integrated spectrum. The PAH emission in the galaxy is
therefore dominated by the diffuse regions, not associated with the
brightest MIR sources.

Regions A, B, and C have derived warm dust temperatures of 187~K
(source A); 179~K (source B); and 201~K (source C), with statistical
uncertainties from the fits of a few K. The lack of higher spatial
resolution photometry beyond 15\micron\ for this system clearly limits
our ability to trace the behavior of colder dust components for
individual regions. The optical depths at 9.7\micron\ of the warm dust
components are 0.76, 0.58, and 0.75 $\pm$ 0.2 for regions A, B, and C,
respectively. In addition, optical depths from the interstellar
radiation field (ISRF, \citet{Mezger82}) are 0.31, 0.47, and 0.24
$\pm$ 0.2 for sources A, B, and C, respectively. Generally speaking,
each region has a similar optical depth at 9.7 \micron, $\sim$0.6-0.8,
and the ISRF optical depth is about uniform for each region,
$\sim$0.3-0.4.

To confirm our estimates for source extinction and age based on MIR
data, an independent assessment using broadband photometry from
\citet{Hecquet87} and a standard extinction curve \citep{Mathis90} was
established. Using the B-R colors published by \citet{Hecquet87} and
the ages derived in our Starburst models (see
\S\ref{sec:StarburstModeling}) we derived E(B-R) for regions A, B, and
C, and compared these values to the standard extinction curve.  The
$\tau$$(9.7)$ values for our three sources are 0.27, 0.17, and 0.16,
respectively.  These values are a factor of two or three less than
those derived from the MIR spectra. The MIR probes more deeply into
dust-enshrouded regions than optical light, which tends to be
dominated by emission from less-obscured stars. Thus MIR indicators of
extinction tend to give higher values than optical.

\subsection{Starburst Modeling}
\label{sec:StarburstModeling}

The atomic lines present in the single source spectra
(Figure~\ref{multispec}) enable us to investigate the nature of the
sources powering the MIR in NGC\,6052. [Ar~{\sc iii}], [Ne~{\sc ii}],
and [S~{\sc iv}] lines are detected in all three regions. By applying
the measurements of these atomic lines to young stellar population
models, we have estimated relative ages for the young stellar
populations of sources A and C, and have established an upper limit
for source B. We used Starburst99 \citep{Leitherer99} to model the
radiation from the underlying stellar population, and MAPPINGS III
\citep{Dopita00, Kewley01} to estimate the flux from the photoionized
regions. The stellar population models were used as input to the
photoionization code. The age of the stellar population was allowed to
vary and the models were run for several metallicities between
Z$_{\odot}$/20 and 2Z$_{\odot}$. The reason we varied the metallicity
was also motivated by the fact that there is a scatter in the
published values of the metallicity of Mrk 297. \citet{Shi05}
estimates [O/H] to be 8.34. Assuming solar abundance of 8.83
\citep{Grevesse98}, this scales to $\sim$ $0.25Z_{\odot}$. Likewise,
\citet{Calzetti94} and \citet{James02} found [O/H] =8.61 and 8.65
respectively (Z $\simeq$ $0.5Z_{\odot}$). Our best fit models are for
solar metallicity. The IMF is fixed at the standard Salpeter values of
$\alpha =$ 2.35, M$_{down} =$ 1.0 M$_{\odot}$ and M$_{up} =$ 100
M$_{\odot}$, and the star formation mode was instantaneous. As such,
the variation of the ionization parameter is only linked to the
variation in age and metallicity of the stellar population. We find
that the young stellar populations in source A is $\sim$3 Myr $\pm$
0.5 Myr, while for source C it is $\sim$5.5 $\pm$ 0.5 Myr. For source
B though we can only place an upper limit of $\lesssim$ 5 Myr due to
the fact that the models were degenerate for region B's atomic line
ratios, which are consistent with ages of both $\sim$1.5 and $\sim$4
Myr.

\section{Discussion}
\label{sec:Discussion}

As mentioned in the previous section and presented in Figure
\ref{multispec}, there is a strong similarity in all three MIR-bright
region spectra in NGC\,6052. Starburst models have shown regions A, B,
and C are nearly coeval, which suggests that some mechanism induced
star formation across the galaxy simultaneously. This mechanism is
most probably the collision of two galaxies as discussed in
\citet{Taniguchi91}, though why the burst of star formation should
come 150 Myr after the two galaxies' closest approach is unclear.

The shallow 9.6\micron\ silicate absorption feature observed in all of
our spectra (indicative of low optical depths), in conjunction with
the morphological properties discussed in \S\ref{sec:GlobalMorphology}
lead us to believe NGC\,6052 has a nonuniform obscuring dust screen
associated with it. The shallow 9.6\micron\ feature implies that the
dominant sources of MIR emission are not embedded dusty star forming
regions but rather exposed in a manner where we can readily sample the
MIR contribution from the photodissociation regions. An embedded
geometry would be proposed if the silicate feature were much deeper,
as is seen in ULIRGs \citep{Armus07}, where hot dust temperatures
suggest that the dust is very close to the hot, young stars heating
it. This is clearly not the case in NGC\,6052. Dust temperatures in
NGC\,6052 of around 200 K (as opposed to the $\sim$700 K found in by
\citet{Armus07} or even the $>$ 300 K dust temperatures implied in the
overlap region of the Antennae described by \citet{Mirabel98}) mean
the dust is relatively far away from the heating source, likely in the
form of a dust screen in front of the exciting sources. A non-embedded
geometry would also explain why the continuum is not as steep as other
embedded systems. However, we believe this screen to be nonuniform
because the system is still in an early stage of interaction and has
not reached the merger phase. There are no apparent symmetries visible
in the MIR morphology of the system; the coincidence of concentrated
MIR flux with a select few HII regions appears arbitrary.

That NGC\,6052 is enshrouded nonuniformly by dust is in general
agreement with other published indications of
nonuniformity. \citet{Rothberg04} found ``large residuals'' in their
K-band model-subtracted image of the galaxy and concluded that
NGC\,6052 has a ``thin diffuse patchy structure''. The predicted
morphology of the system based on dynamical modeling of the stellar
component by \citet{Taniguchi91} demonstrated that when they can
reproduce the overall morphology of the system (see their Figures 2
and 4), one of the tidal tails extends to the north of the central
bulge of the galaxy, across the location of regions A and C, while the
other tidal tail extends eastward, across region B and including
region D. Based on this scenario it is thus conceivable that clumpy
gas and dust associated with these tails are associated with the four
distinct regions we see in the infrared.

In \S\ref{sec:GlobalMorphology} we noted that region A, which emits
most at 16 \micron, contributes 18.5 \% of the total flux. This fact
begs comparison to the fluxes of the obscured regions of the Antennae
galaxies and IC\,694, since both of these systems have off-nuclear
sources which contribute largely to the MIR emission. The hidden
source in the Antennae galaxy contributes 15 \% of the total
luminosity between 12.5\micron\ and 18\micron\ \citep{Mirabel98}, and
IC\,694 contributes $\sim$ 26 \% of the total flux between 12\micron
and 18\micron. However, neither of the hidden sources present such
spectral similarities with other regions of their host galaxy as
region A does to the other regions of NGC\,6052 (cf. the spectra
presented in \citet{Vigroux96,Mirabel98,Gallais04}). Also, we now know
that all of the MIR-bright regions in NGC\,6052 are optically visible
(see Table~\ref{srcid}). For this reason, and because there is
currently no evidence which allows region A to be considered an AGN
(as is the case with the obscured source in IC\,694), a comparison
between NGC\,6052 and these two objects yields very little information
about the physical properties of the MIR-bright regions.

Because all of the MIR bright regions in NGC\,6052 lie within the dust
trail or in one of the two tidal tails, we compare NGC\,6052 to other
galaxies with bright sources existing in the tidal tails. One obvious
comparison can be drawn with the MIR-bright region in NGC\,2207 /
IC\,2163. This region (called ``feature i'') contributes 12\% of the
total 24\micron flux, and also dominates the H-$\alpha$ and radio
continuum emission \citep{Elmegreen06}. The implication of this
finding is that ``feature i'' is a compact region of star
formation. Our region A also exhibits bright MIR flux and
corresponding radio emission (see \S\ref{sec:RadioContinuumEmission}
and \citet{Yin94}). Tidal tail-induced star formation is also present
in the Arp\,82 system, where the brightest off-nuclear clump
contributes $\sim$ 7\% of the total 24 \micron\ flux
\citep{Hancock07}. Region D likewise contributes $\sim$ 5\% of the
total 16 \micron\ flux.

We can also calculate the MIR luminosities of the regions in NGC\,6052
and compare them with studies which include other interacting galaxy
samples \citep[i.e.][]{Smith07}. For regions A, B and C we find that
the 3.6\micron\ luminosities, which we treat as a proxy of the stellar
mass content, are 1.34, 1.38 and 1.77$\times$10$^{42}$ ergs s$^{-1}$
respectively. The 8.0\micron\ luminosities, which probe the emission
from PAH molecules and thus are tracers of the star formation
activity, are 7.46, 7.13 and 8.30$\times$10$^{42}$ ergs s$^{-1}$ for
the same three regions. These values are very similar to values found
in the spiral galaxy and the tidal feature samples of \citet{Smith07}
(see their Figs. 12 and 13) if one takes into account that there is an
RMS of $\sim$0.5 in dex in the values of the Smith et al. sample. When
we examine region D in the tidal tail of NGC6052 we find that its 3.6
and 8.0\micron\ luminosities are 7.05$\times$10$^{41}$ ergs s$^{-1}$
and 2.50$\times$10$^{42}$ ergs s$^{-1}$ respectively. These
luminosities are slightly higher than the mean value of the tidal
features in the Smith et al. sample. However, NGC\,6052 is also nearly
3 times more distant than the furthest source in the Smith et
al. survey, so in our photometry for all four regions of our galaxy we
may be including other parts of the galaxy which would have likely
been excluded by \citet{Smith07}. This inherent error is likely
biasing our luminosity measurements towards measurements made of
entire galaxies, and away from small-scale tidal features.

\section{Conclusions}
\label{sec:Conclusion}

We have presented photometric and spectroscopic results of the
NGC\,6052 system. The main conclusions of our analysis are as follow:
(1) The brighteset MIR region, labeled A, contributes $\sim$ 18.5\% of
the total 16\micron\ flux, and resembles the brightest 24 \micron\
source from NGC\,2207 / IC\,2163 photometrically, but is dissimilar
spectroscopically from the hidden sources in the Antennae galaxies and
Arp\,299; (2) the spectra of regions A, B, and C are all surprisingly
similar, and are consistent with dust emission from nearly coeval
young stellar populations; (3) the geometry of this system includes a
patchy dust screen of colder material in front of the young stellar
clusters, and these clusters are not deeply embedded in dusty
material; and (4) region D has MIR properties which are typical of
those found in tidal tails.

\acknowledgments

This work is based [in part] on observations made with the Spitzer
Space Telescope, which is operated by the Jet Propulsion Laboratory,
California Institute of Technology, under NASA contract 1407. Support
for this work was provided by NASA through Contract Number 1257184
issued by JPL/Caltech.

D.G. Whelan would like to thank an anonymous referee whose comments
greatly improved this paper, and H. Spoon for stimulating
conversations. We would also like to thank C. Markwardt for
distributing the MPFIT
package\\(\url{http://cow.physics.wisc.edu/craigm/idl/mpfittut.html}).

Some of the data presented in this paper were obtained from the
Multimission Archive at the Space Telescope Science Institute (MAST).
STScI is operated by the Association of Universities for Research in
Astronomy, Inc., under NASA contract NAS5-26555. Support for MAST for
non-HST data is provided by the NASA Office of Space Science via grant
NAG5-7584 and by other grants and contracts.

This research has made use of the NASA/ IPAC Infrared Science Archive,
which is operated by the Jet Propulsion Laboratory, California
Institute of Technology, under contract with the National Aeronautics
and Space Administration.

{\it Facilities:} Spitzer (IRS)

\clearpage

\begin{deluxetable}{ccccccccccc}
  \tabletypesize{\tiny} \setlength{\tabcolsep}{0.02in}
  \tablecaption{Source Identification\tablenotemark{a}\label{srcid}}
  \tablewidth{0pt}
  \tablehead{\multicolumn{3}{c}{Labels} & \multicolumn{8}{c}{Fluxes in mJy} \\ \colhead{Spitzer} & \colhead{Metcalfe\tablenotemark{b}} & \colhead{Hecquet\tablenotemark{c}} & \colhead{J band} & \colhead{H band} & \colhead{K band} & \colhead{3.6$\micron$} & \colhead{4.5$\micron$} & \colhead{5.8$\micron$} & \colhead{8.0$\micron$} & \colhead{16.0$\micron$}}
  \startdata
  A   & 1    & 7,8   & 0.96                   & 1.16                   & 1.05                   & 2.68 & 2.46 & 10.6 & 33.1 & 48.0 \\
  B   & 2    & 22,23 & 1.89                   & 2.42                   & 2.11                   & 2.76 & 2.21 & 9.73 & 31.6 & 36.1 \\
  C   & 4    & 15    & 1.26                   & 1.43                   & 1.34                   & 3.53 & 2.42 & 11.5 & 36.8 & 30.1 \\
  D   & 7    & 41    & 1.21                   & 1.36                   & 1.26                   & 1.41 & 1.04 & 3.68 & 11.1 & 11.9 \\
  all & $--$ & $--$  &43.65\tablenotemark{d}  &56.23\tablenotemark{d}  &38.90\tablenotemark{d}  &22.3  &16.4  &63.2  &179   &260   \\
  \enddata
\tablenotetext{a}{Errors in flux densities are measured to be about 12 \% systematically.}
\tablenotetext{b}{Taken from \citet{Metcalfe05} Figure 1, sources 1, 2, 4, amd 7 refer to the MIR regions they labeled from the ISOCAM 14.3 \micron\ image}
\tablenotetext{c}{Taken from \citet{Hecquet87} Figure 2b, a B-R color image of NGC\,6052}
\tablenotetext{d}{Global flux density values for the J, H, and K bands are taken from \citet{Spinoglio95}}
\end{deluxetable}

\clearpage

\begin{deluxetable}{lcccc}
  \tabletypesize{\small} \setlength{\tabcolsep}{0.02in}
  \tablecaption{Mid-IR Emission Features of NGC\,6052\label{tabflux}}
  \tablewidth{0pt}
  \tablehead{\colhead{} & \multicolumn{4}{c}{Flux $\times 10^{-20}$\,W\,cm$^{-2}$}\\
    \colhead{Feature} & \colhead{Region A} & \colhead{Region B} & \colhead{Region C } & \colhead{Total}}
\startdata
  PAH 6.2\micron\                    & 4.32  $\pm$0.04   & 3.41 $\pm$0.03  & 4.92 $\pm$0.04  &  65.6$\pm$0.8 \\
  PAH 7.7\micron\                    &15.8   $\pm$0.2    &15.0  $\pm$0.7   & 1.85 $\pm$0.03  & 198  $\pm$6  \\
  PAH 8.6 \micron\                   & 2.93  $\pm$0.5    & 2.96 $\pm$0.6   & 3.45 $\pm$0.06  &  30.1$\pm$0.7 \\
  PAH 11.3\micron\                   & 3.72  $\pm$0.8    & 3.4  $\pm$0.1   & 4.4  $\pm$0.1   &  38.9$\pm$0.9 \\
  PAH 12.7\micron\                   & 2.75  $\pm$0.4    & 2.56 $\pm$0.4   & 2.36 $\pm$0.08  &  15.0$\pm$0.5 \\
  ${\rm[ArIII]}\lambda$8.99\micron\  & 0.139 $\pm$0.002  & 0.196$\pm$0.001 & 0.071$\pm$0.002 &   5.7$\pm$0.3 \\
  ${\rm[SIV]}\lambda$10.51\micron\   & 0.43  $\pm$0.02   & 0.41 $\pm$0.04  & 0.20 $\pm$0.02  &   2.3$\pm$0.5 \\
  ${\rm[SIII]}\lambda$19.01\micron\  & $---$             & $---$           & $---$           &   4.4$\pm$0.5 \\
  ${\rm[NeII]}\lambda$12.81\micron\  & 0.65  $\pm$0.11   & 0.95 $\pm$0.11  & 0.74 $\pm$0.11  &   6.3$\pm$0.6 \\
  ${\rm[NeIII]}\lambda$15.80\micron\ & $---$             & $---$           & $---$           &   5.3$\pm$0.5 \\
  $H_{2}S(3)$$\lambda9.7\mu$m        & 0.0792$\pm$0.007  & 0.171$\pm$0.01  & 0.17 $\pm$0.02  &   8.9$\pm$0.7 \\
  $H_{2}S(2)$$\lambda12.3\mu$m       & 0.10  $\pm$0.01   & 0.16 $\pm$0.01  & 0.101$\pm$0.01  &   6.5$\pm$0.3 \\
\enddata
\end{deluxetable}

\clearpage

\clearpage

\clearpage

\clearpage

\end{document}